\documentclass[aps,notitlepage,a4paper,superscriptaddress,11pt]{article}

\usepackage{multicol}
\usepackage[utf8]{inputenc}
\usepackage[dvips]{color}
\usepackage[toc]{appendix}
\usepackage[hang,flushmargin]{footmisc} 
\usepackage{titlesec,titletoc,ntheorem-hyper,authblk,abstract,latexsym,microtype,booktabs,amsmath,hyperref,cleveref,fancyhdr,wrapfig,array,amsfonts, amssymb,geometry,appendix,cite, secdot, tensor}
\usepackage{xcolor}

\newcommand{\be}{\begin{equation}}
\newcommand{\ee}{\end{equation}}
\newcommand{\bea}{\begin{eqnarray}}
\newcommand{\eea}{\end{eqnarray}}
\newcommand\blfootnote[1]{%
  \begingroup
  \renewcommand\thefootnote{}\footnote{#1}%
  \addtocounter{footnote}{-1}%
  \endgroup
}

\usepackage{graphicx,lipsum, cuted}
\usepackage{caption}
\usepackage{subfigure}
\usepackage{amsmath}
\usepackage{amsfonts}
\usepackage{amssymb}
\usepackage{stackengine}
\usepackage{mathtools}
\numberwithin{equation}{section}
\usepackage{titlesec}
\usepackage{sectsty}
\sectionfont{\bfseries\Large\raggedright}

\numberwithin{subcase}{case}
\usepackage{framed}
\allowdisplaybreaks
\usepackage[nottoc]{tocbibind}
\usepackage[free-standing-units]{siunitx}
\usepackage{helvet}
\usepackage{url}
\usepackage{titletoc}
\usepackage{blindtext}
\usepackage[gen]{eurosym}

\title{Artificial Hawking radiation, weak pseudo-Hermiticity and Weyl semimetal blackhole analogy}
\author{Bijan Bagchi and Sauvik Sen}
 \affil{Department of Physics, School of Natural Sciences,\\ Shiv Nadar University,\\ Gautam Buddha Nagar, Greater Noida,\\
Uttar Pradesh 203207, India}
\date{\today}

\begin{document}

\maketitle

\begin{abstract}
    We examine the possibility of artificial Hawking radiation by proposing a non-$\mathcal{PT}$-symmetric weakly pseudo-Hermitian two-band model containing a tilting parameter by pursuing Weyl semimetal blackhole analogy. We determine the tunneling probability using such a Hamiltonian through the event horizon that acts as a classically forbidden barrier. 
    \end{abstract}

\blfootnote{E-mails: bbagchi123@gmail.com, sauviksen.physics@gmail.com}


\section{Introduction}

Non-Hermitian quantum mechanics is an emerging field of interest with a wide range of applications\cite{ash, moi}. In particular, the sub-class embodying $\mathcal{PT}$-symmetry has proved to be an area of continuous activity\cite{ben, ram}. In fact, over the past two decades a large family
of exactly solvable $\mathcal{P}\mathcal{T}$-symmetric systems has been discovered reflecting their intriguing spectral properties.
Briefly, $\mathcal{PT}$-symmetry addresses a complex extension of quantum mechanics that is controlled by the combined actions of parity ($\mathcal{P}$) and time reversal ($\mathcal{T}$) transformations\cite{ben} namely, $\mathcal{P}: x \rightarrow -x, \quad p \rightarrow - p, \quad \mathcal{T}: x \rightarrow x, \quad p \rightarrow - p, \quad i \rightarrow -i$.
Non-Hermitian Hamiltonians undergo non-unitary evolution and generally describe open systems through gain and loss of particles. 

Hamiltonians respecting $\mathcal{PT}$-symmetry may exhibit, under certain condition related to $\mathcal{PT}$ being exact, real spectra of eigenvalues, implying balanced loss and gain. However, $\mathcal{PT}$-symmetry is neither necessary nor sufficient for the reality of the spectrum. 
An exceptional point (EP) appears where symmetry breaking occurs\cite{heiss, ply}. In such a situation one finds the eigenvalues corresponding to two states to coalesce and the accompanying eigenfunctions become linearly dependent with respect to each other. However, in approaching the EP, the phases of the eigenfunctions may not show robustness, as a consequence information from outside may leak into the system\cite{rot}. EPs 
play an important role in the characterization of non-Hermitian Hamiltonians. For a recent survey of the existence of an EPs see \cite{fmf, ssen, serg}.

The idea of $\mathcal{PT}$-symmetry has found extension in the formulation of pseudo-Hermiticity. For the pseudo-Hermitian operators one takes recourse to the concept bi-orthogonality of wavefunctions\cite{mos1, mos2}. The Hamiltonian $H$ is called pseudo-Hermitian if there exists a Hermitian invertible operator $\eta$ satisfying

\begin{equation}
H^{\dagger}=\eta H \eta^{-1}
\end{equation}
where the Hermitian conjugation is taken in the Hilbert space that is endowed with a specific inner product. A subtle point concerning the role of (1.1) in non-Hermitian systems may be made here. Consider a simple $2 \times 2$ matrix model with $\eta =\sigma_z$. We recognize $\eta$ to be Hermitian and indefinite in character, i.e. with eigenvalues $\pm 1$. This implies that $\eta$ can act as an indefinite metric in a Krein space, as elaborated in \cite{gue1, gue2}, so that H itself is self-adjoint in such a Krein space. But operators self-adjoint in Krein spaces are known to have a spectrum which, in general, is symmetric with regard to the real axis in the complex spectral plane. This means that such an H can have pairwise complex conjugate eigenvalues as well as EP-type degenerate eigenvalues.
Concerning the discussion in \cite{mos2},  the reality of the spectrum of H holds not for general (indefinite) $\eta$ but only for the class of $\eta_+$ which are assumed as
positive definite Hilbert space metrics. In the literature, this difference is emphasized by referring this subclass of pseudo-Hermitian H with $\eta=\eta_+$ as quasi-Hermitian ones.

Like for unbroken $\mathcal{PT}$ systems,  pseudo-Hermitian systems can be constructed to encounter full balance of loss and gain (see, for example, \cite{luo} and references therein).  However, in what follows, we will focus on a weak pseudo-Hermitian operator $\eta$ that is not restricted to be Hermitian\cite{solm}. Such a relaxation opens up possibility of connecting to a wider class of non-Hermitian systems\cite{bag1, ahm, zno1, mos3}.

Lately, much interest has been focused on the issue of
phases that are special to non-Hermitian systems
and do not appear in the Hermitian setups\cite{berg, xu1, xu2, chr}. In particular,  $\mathcal{P}\mathcal{T}$-symmetry is observed to have a subtle role to play in stable nodal points for gapped and gapless semimetals\cite{goe, arm} where Bloch bands constitute invariants. Of course, non-Hermitian support for stable phases has been in the news for sometime\cite{wei, yuce}. That band crossing is prevalent in three-dimensional systems is known for sometime. Because of the role of $\mathcal{PT}$-symmetry, stable nodal points may occur in lesser dimensions \cite{ber}.

A point was made a few years ago about the question of whether real black holes can emit Hawking radiation and whether non-trivial information can be gathered about about Planckian physics \cite{unr}. Very recently, De Beule et al \cite{chr} (see also \cite{sab}) have made an explicit analysis of the existence of artificial event horizon
in Weyl semimetal heterostructures. The electronic analogs of stimulated Hawking
emission was studied and physical observables were addressed in the context of Weyl semimetal black hole analogs. A related work  \cite{sta} looked at the $\mathcal{PT}$ symmetry-protected cones and analogue Hawking radiation was explored.

The aim of this note is to propose a weak pseudo-Hermitian two-band model containing a tilting parameter that reveals the signature of an artificial Hawking radiation. Towards this end, we review briefly in section 2 the background of the two-band structure where we justify how an artificial Hawking radiation can plausibly take place. In section 3, we calculate the contribution of such a Hamiltonian to the tunneling. Finally, in section 4, we present a summary.

\section{Pseudo-Hermitian Hamiltonian}

To begin with, we write down a tilted Weyl Hamiltonian distorted in the x-direction \cite{sta}

\begin{equation}
    H = \xi p_x \P + \vec{p} \cdot \vec{\sigma}
\end{equation}
where $\xi \in \Re$ is the tilting parameter, $\P$ is the three-dimensional identity matrix, $\vec{p}$ is the three-dimensional momentum with components $(p_x, p_y, p_z)$, $\vec{\sigma}$'s are a set of Pauli matrices $(\sigma^x, \sigma^y, \sigma^z)$ which are Hermitian and unitary and set the Fermi velocity to be 1. The accompanying energy eigenvalues can be readily evaluated

\begin{equation}
    E_{\pm} = \xi p_x \pm \sqrt{p_x^2 + p_y^2 +p_z^2}
\end{equation}
where the two signs reflect two zones of a cone, the upper and lower, in the energy-momentum space. Weyl semimetals with tilted nodes admit black and white hole analogs \cite{sab,vol1}.  Experimental discovery of Weyl and
Dirac semimetals (see, for example, \cite{bur1}) has led to intense study of the Weyl Hamiltonian \cite{bur2, chr, sab}. The study of H reveals that the Weyl cones touch when the two energies become equal and may cross the Fermi level if overtilting happens ($|\xi| > 1$). The existence of strongly tilted Weyl cones has been proposed to exist in layered transition metals\cite{arm}. 

Adopting the standard tetrad representation \cite{vol1, chr, sta}

\begin{equation}
    H = \tensor{e}{^i_a} p_i\sigma^a+\tensor{e}{^i_0} p_i\P
\end{equation}
where $e^\mu_\alpha$ are the vielbiens satisfying the orthonormality condition $\tensor{e}{^a_\mu} \tensor{e}{^\mu_b}=\tensor{\delta}{^{b}_{a}}$, $\mu,\alpha= (0,x,y,z)$ and $i,a= (x,y,z)$ subject to the inner
product-signature constraint 
\begin{equation}
    g^{\mu\nu}=\tensor{e}{^\mu_\alpha} \tensor{e}{^\nu_\alpha} \eta^{\alpha \beta}
\end{equation}
where $ \eta^{\alpha \beta} = diag(-1,1,1,1)$ is the Minkowski metric
of flat spacetime, comparison with $(2.1)$ yields for the line element 

\begin{equation}
    ds^2 = g_{\mu\nu} dx^{\mu} dx^{\nu} = -(1-\xi^2) dt^2 + 2\xi dx dt + dx^2
\end{equation}
$|\xi| = 1$ represents the event horizon.
Through critical-tilting of the cone, a phase transition takes place and emission of  Hawking-like radiation can be envisaged due to pair production that involve pairs of light-like particles and their anti-particles. As remarked by Carlip \cite{car} for a black hole that is initially
formed as a result of collapse of matter in a pure quantum state and subsequently evaporating completely into Hawking radiation, a transition is represented from a pure state to a mixed state thereby violating unitarity, a situation unfamiliar in standard quantum mechanics. However, if the Hawking radiation is involved in a pure state, this would appear to
require correlations are implied between the early and late Hawking particles that otherwise were not in causal contact (see \cite{alm}). Note that Hawking radiation\cite{haw} refers to thermal radiation emitted by a black hole off its event horizon if the quantum effects are taken into account. The contention is that pair production leads to one of the particles escaping the boundary of the black hole to infinite space leaving the other of negative energy returning into it. Incessant transitions of negative energy particles back into the black hole inevitably reduces its mass until the whole black hole disappears into a cloud of radiation. In this way, an analogy has been drawn between Weyl semimetals with inhomogeneous tilting and spacetime conforming to black holes triggering off the idea of an artificial Hawking radiation in Weyl semimetals\cite{vol1, vol2, sol, huh}. 

Consider the following non-Hermitian Hamiltonian of a topologically insulated two- band model
\begin{equation}
    \mathcal{H} = p_x\sigma^x + p_y\sigma^y + \iota (p_z - \lambda p_x)\sigma^z, \quad \lambda \in R
\end{equation}
where the third term in the right specifies the presence of non-Hermiticity that contains tilting in the x-direction, with $\lambda$ signifies the coupling strength. One can easily check that $\mathcal{H}$ does not commute with the $\mathcal{PT}$ operator. Actually, the explicit presence of the $p_y$ term in (2.6) spoils the $\mathcal{PT}$ character of $\mathcal{H}$. 

For $\mathcal{H}$, the corresponding eigenvalue problem gives 

\begin{equation}
 |\mathcal{H} -\kappa I| = 0 \implies   \kappa^2 = p_x^2 + p_y^2 - (p_z -\lambda p_x)^2
\end{equation}
implying the existence of a pair of eigenvalues

\begin{equation}
\kappa_{\pm} = \pm\sqrt{(p_x^2+p_y^2)-(p_z-\lambda p_x)^2}
\end{equation}
with the associated eigenvectors

\begin{equation}
\left [\frac{\iota(p_z-\lambda p_x)+\sqrt{(p_x^2+p_y^2)-(p_z-\lambda p_x)^2}}{p_x + \iota p_y},1 \right ]^T, \quad \left [\frac{\iota(p_z-\lambda p_x)-\sqrt{(p_x^2+p_y^2)-(p_z-\lambda p_x)^2}}{p_x + \iota p_y},1 \right]^T
\end{equation}
where T stands for the transpose. It is readily seen from (2.8) that for the condition $\lambda_{+} = \frac{p_z + \sqrt{p_x^2+p_y^2}}{p_x}$ the eigenvalues $\kappa_+$ and $\kappa_-$ coincide to become vanishing and the associated eigenvectors coalesce to the form $\left [-\frac{\iota \sqrt{p_x^2 + p_y^2}}{p_x + \iota p_y},1 \right ]^T$ (similarly, for $\lambda_{-} = \frac{p_z - \sqrt{p_x^2+p_y^2}}{p_x}$, the associated eigenvectors coalesce to $\left [\frac{\iota \sqrt{p_x^2 + p_y^2}}{p_x + \iota p_y},1 \right ]^T$). Thus $\kappa_1 = \kappa_2 = 0$ could be identified as the EP and passing through it is accompanied by a spontaneous symmetry breaking of eigenstates and a change of the eigenvalues implying a crossover from their real character to a pair of complex entities. At the EP we have the relation

\begin{equation}
    p^{\pm}_z=\lambda p_x \pm \sqrt{p_x^2 + p_y^2}
\end{equation}

On comparing with its Hermitian counterpart of $\mathcal{H}$, namely

\begin{equation}
    H = \lambda p_x \P + p_x \sigma^x + p_y \sigma^y
\end{equation}
whose energy eigenvalues read
\begin{equation}
    E_{\pm} = \lambda p_x \pm \sqrt{p_x^2 + p_y^2}
\end{equation}
one easily observes that $E_{\pm}$ and $p^{\pm}_z$ are interchangeable entities affording $p^{\pm}_z$ to be interpreted as a Hamiltonian-like operator. 

We offer an interesting interpretation here. Although $\mathcal{H}$ is not entirely $\mathcal{PT}$-symmetric by itself, we can treat it as a combination of two Hamiltonians one of which is Hermitian ($\mathcal{H}_h$) while the other is weak pseudo-Hermtian ($\mathcal{H}_w$) with respect to $\mathcal{PT}$. Interfacing the Hermitian and non-Hermitian systems has been a topic of interest in the literature\cite{ben3, zno2}. Thus, we write 

\begin{equation}
\mathcal{H} = \mathcal{H}_h + \mathcal{H}_{w}
\end{equation}
where 

\begin{eqnarray}
&& \mathcal{H}_h = p_y \sigma^y\\
&& \mathcal{H}_{w}= p_x\sigma^x + \iota (p_z - \lambda p_x)\sigma^z
\end{eqnarray}
While the Hermiticity of $\mathcal{H}_h$ is trivial, the weak pseudo-Hermiticity of $\mathcal{H}_{w}$ follows from the condition

\begin{equation}
    \mathcal{H}^{\dagger}_{w}= \rho \mathcal{H}_{w} \rho^{-1}
\end{equation}
as is evident on employing $\rho =-\iota\sigma^x (\neq \rho^\dagger)$. However, it should be borne in mind that $\mathcal{H}_{w}$ is not $\rho$-symmetric. The operator $\rho$ is not invariant under the conventional transformation of $\mathcal{PT}$. Examples of Hamiltonians endowed with the property of weak pseudo-Hermiticity but not being $\mathcal{PT}$-symmetric have been explicitly constructed before \cite{bag1}. 

%In contrast to the weak pseudo-Hermiticity as just now defined, 
It needs to be emphasized that in the scheme of \cite{sta}, the idea of $\mathcal{PT}$-symmetry preservation was pursued. The latter was defined in terms of an operator $\mathcal{P}$ and a non-conventional $\mathcal{K}$-matrix such that $\mathcal{KK}^* = 1$, $\mathcal{P}^2 = 1$, and  $\mathcal{KP}^* = \mathcal{PK}$. At first sight, it appears that
such a model differs in spirit from the present work. However, a little observation reveals that this is not so. In fact,
%What concerns the comment four lines below (2.16) so the authors are strongly encouraged to do a most explicit comparison of the rather special PT symmetry condition in [32] with their own weak pseudo-Hermiticity condition.
using standard Pauli matrix representations for $\mathcal{P}$ and $\mathcal{K}$ namely, $P=\sigma_y, K=\sigma_z$ which imply $\mathcal{PK}=i\sigma_x$, it turns out that $H_{\mathcal{PK}}=\sigma_x H_{\mathcal{PK}}^* \sigma_x.$ Resolving this matrix constraint generates a general structure
$H_{\mathcal{PT}} = a_1 I_2 +ia_2 \sigma_z +b_1 \sigma_x -b_2 \sigma_y$, where the coefficients $a_r$ and $ b_r$, with $r = 1, 2$ are constants.
Obviously, this $H_{\mathcal{PK}}$ appearing in [32] structurally coincides with our (2.13) - (2.15).
%\sigma_x={{0,1},{1,0}}, \sigma_y={{0.-i},{i,0}}, \sigma_z={{1,0},{0,-1}} in 
%the definitions of [32] yields for the matrices defined in (17), (18) of [32]

%P=\sigma_y, K=\sigma_z

%so that for PK in (18) it holds

%PK=i\sigma_x

%and, therefore, (18) boils down to

%H_{PT}=\sigma_x H_{PT}^* \sigma_x.

%Resolving this matrix constraint yields for the H_{PT} in [32] a general structure

%H={{a_1+ia_2, b_1+ib_2},{b_1-ib_2, a_1-ia_2}}
%= a_1 I_2 +ia_2 \sigma_z +b_1 \sigma_x -b_2 \sigma_y.

%Obviously, this H_{PT} in (19) of [32] structurally coincides with (2.13) - (2.15) in the authors present manuscript.

Focusing on $\mathcal{H}_{w}$, the eigenvalue problem $|\mathcal{H}_{w}-\kappa I|=0$ transforms to 
\begin{equation}
 \kappa^2 = p_x^2 - (p_z - \lambda p_x)^2 \end{equation}
showing that the eigenvalues of $\mathcal{H}_{w}$ are $\pm\sqrt{p_x^2-(p_z-\lambda p_x)^2}$ with the corresponding eigenvectors

\begin{equation}
\left [\frac{\iota(p_z-\lambda p_x)-\sqrt{p_x^2-(p_z-\lambda p_x)^2}}{p_x + \iota p_y},1 \right]^T, \quad \left [\frac{\iota(p_z-\lambda p_x)+\sqrt{p_x^2-(p_z-\lambda p_x)^2}}{p_x + \iota p_y},1 \right ]^T
\end{equation}
The eigenvalues $\kappa_+$ and $\kappa_-$ both collapse to zero-value when the condition $\lambda_{+} = \frac{p_z}{p_x} + 1$ is imposed. The system thus has an EP here where the eigenvectors coalesce to the form $\left[-\iota, 1 \right ]^T$ (for $\lambda_{-} = \frac{p_z}{p_x} - 1$, the eigenvectors coalesce to $\left[\iota, 1 \right ]^T$). At the EP we obtain the underlying result

\begin{equation}
    p^{\pm}_z = p_x \pm \lambda |p_x|
\end{equation}

For $\mathcal{H}_{w}$, the corresponding Hermitian Hamiltonian is

\begin{equation}
    H_h = p_x\sigma^x + \lambda p_x \P
\end{equation}
which supports the energy eigenvalues 

\begin{equation}
    \mathcal{E}_{\pm} = p_x \pm \lambda |p_x| 
\end{equation}
The forms of $\mathcal{E}_{\pm}$ and $ p^{\pm}_z$ are similar. 
One checks that (2.16) and (2.18) are respectively reduced versions of (2.8) and (2.10) when $p_y$ is absent.

Employing the same arguments leading to (2.5), here too with the Hamiltonian $H_h$ results in a similar metric containing the coupling parameter $\lambda$ and with an additional presence of a term $dy^2$. If we consider a slice of $y = \mbox{constant}$, then the latter contribution drops out and the metric transforms to  the Schwarzschild black hole in Painlev\'{e}-Gullstrand
coordinates\cite{pain, gul, sta}

\begin{equation}
    ds^2 = -\left (1-\frac{2\mathcal{M}}{r}\right ) d\tau^2 + 2 \sqrt{\frac{2\mathcal{M}}{r}} dr d\tau + dr^2 
\end{equation}
with evident identifications of x with r, and t with $\tau$, the Painlev\'{e} time and fixing $\lambda$ as the quantity $\sqrt{\frac{2\mathcal{M}}{r}}$, $\mathcal{M}$ denoting the mass of the black hole. In arriving at (2.19) we used the tetrad formalism for the Hermitian Hamiltonian $H_h$. We now turn to the process of tunneling across the black hole horizon.\\

\section{Tunneling probability}

The tunneling probability of the Hawking radiation is simple to compute \cite{bag2, ali, jur, sta}. The contribution comes from the first and third terms of the right side of $\mathcal{H}$. In our case, we evaluate the tunneling from the weak pseudo-Hermitian part (2.13) which is our guiding Hamiltonian. In fact, the y-component cannot be present in $\mathcal{H}_{w}$ to preserve its weak pseudo-Hermiticity. 
%the EP related feature demonstrated in the earlier section Crucial for the physics is the fact that the Hamiltonians of both models structurally coincide and that the Hawking radiation aspects depend only on this specific matrix structure and the EP related features.

The particle escaping from the black hole has an energy $\omega$ and so the mass of black hole is reduced from  from $\mathcal{M}$ to $\mathcal{M}-\omega$. 
More precisely, when the pair production is happening inside the event horizon, the positive energy particle will tunnel out\cite{par, fle, bag2}. The region between $r_{in}$ (radius before the emission of the particle) and $r_{out}$ (radius after the emission of the particle) that separates the two points serves as a possible barrier for the tunneling particle to go across. The particle created during pair creation would have an energy below the acting barrier, thereby replicating a classically prohibited zone. Since the action in this region is imaginary, we may effectively estimate the tunnelling probability using the semiclassical WKB approximation.

For our calculation of the tunneling corresponding to the weak pseudo-Hermitian component $\mathcal{H}_{w}$ of $\mathcal{H}$, the procedure is standard as elucidated in \cite{par}.
The situation that we encounter resembles a contrived scenario of Hawking radiation, where the pair production of particles occurs near the event horizon of the black hole, which is given by the metric $(2.22)$, the event horizon
playing a potential barrier for the outgoing particle. Using the Legendre transformation, the associated Lagrangian $\mathcal{L}_{w}$ for $\mathcal{H}_{w}$ can be expressed as

\begin{equation}
    \mathcal{L}_{w} = \hspace{2mm} p_r \cdot \dot{r}  - \mathcal{H}_{w}
\end{equation}

We focus on the condition of $\kappa=0$ at the EP. From (2.8) which then reads  $p_x^2-(p_z-\lambda p_x)^2=0$ in the absence of $p_y$ we get, after carrying out the identifications mentioned when dealing with (2.22), the following expression 

\begin{equation}
    p_r= -\frac{\sqrt{2\mathcal{M}r}p_z}{r-2\mathcal{M}} - \frac{r p_z}{r-2\mathcal{M}}
\end{equation}
where we have chosen the negative sign for the outgoing particle. Using (3.1) and considering expansion in the neighbourhood of $r = 2M$, the imaginary part of the action ($\zeta=\int \mathcal{L}_{w}dt$) appearing in the tunneling probability 
$\Gamma \sim e^{-2 Im(\zeta)}$  is straightforwardly given by

\begin{equation}
    Im \zeta = Im \int_{2\mathcal{M}}^{2(\mathcal{M}-\omega)}  \hspace{2mm} \left (\displaystyle\dfrac{- \sqrt{2\mathcal{M}r}p_z - r p_z}{r-2\mathcal{M}} \right) \cdot dr 
\end{equation}
 where the contribution of the second term from (3.1) is dropped because at the exceptional point where the two eigenstates coalesce, the time taken for the transition is rather small and hence the neglect of the integral over dt is justified \cite{heiss,ben1}. Indeed as pointed out by Heiss \cite{heiss}, the EP, serving as a critical point, could be interpreted as a turning point and further that for the level splitting by a potential barrier, the feature of quantum tunneling is associated with the coalescence of two levels at an appropriate value of the barrier strength.

To determine the integral in (3.3) we make the substituion of $r - 2M = r_*$ so that $dr = dr_*$ so that (3.3) is re-expressible as

\begin{equation}
    Im \zeta = Im \int_{0}^{-2\omega}  \hspace{2mm} \left (-\displaystyle \frac{\sqrt{2\mathcal{M}(2\mathcal{M}+r_*)}}{r_*}p_z - \frac{(2\mathcal{M}+r_*)}{r_*} p_z \right ) dr_*  
\end{equation}

Evaluation of this integral in (3.4) can be done by Taylor expansion of the integrand  and concentrating on the pole around $r = 0$. Using the Cauchy integral formula $\int_C \frac{dz}{z} dz = 2\pi \iota$, we obtain the value $8\pi M p_z$ so that the particle-contribution to the tunneling probability $\Gamma$ is given by the following result 

\begin{eqnarray}
    \Gamma \sim e^{-2Im(\zeta)} = e^{-16 \pi \mathcal{M} p_z}
\end{eqnarray}
We want to point out that the expression for tunneling probability is not unique and quite often variations from the Parikh-Wilczek estimate \cite{par} of $8\pi\mathcal{M}$ are noticed in the literature (see, for example, \cite{bag2}). The effects of rotation can also have an impact on the tunneling probability \cite{wu}.  When compared with the work of \cite{par}, we find that the argument of the exponential in (3.5) has a value which is a factor of 2 too large. On inspection we find that from (3.5) that the exponential can be compared with the Boltzmann factor $(e^{-\frac{\omega}{T_H}})$ for a particle with energy $\omega$ that is getting emitted. Here $T_H$ is the temperature of the corresponding Hawking radiation. We remark here that although the structural nature of our model and that of \cite{sta} coincides, the feature of Hawking radiation, including the estimate in (3.5), drawn from our analysis of the location of EPs, differs from it.

It is relevant to mention that experimental studies of artificial Hawking radiation in ultra-soft laser pulse filaments have been carried out in \cite{bel} by creating an effective flowing medium that mimics certain characteristic features of black hole physics. Indeed analogous black hole horizons may be duplicated with photon emissions similar to Hawking radiation \cite{fac}.

\section{Summary}

Against the background of some recent works on artificial event horizons, 
we have set up in this paper, a Hamiltonian relevant for a tilted two band model that goes with the spirit of modeling a Weyl semimetal. Our proposed Hamiltonian is weakly pseudo-Hermitian rather than $\mathcal{PT}$-symmetric. The notion of weak pseudo-Hermiticity extends the definition of the underlying operators to exclude the constraint of the operator to be Hermitian. However, our Hamiltonian is not $\mathcal{PT}$-symmetric.
Moreover, our scheme reflects similar features of a Schwarzschild black hole when translated to the Painl\'{e}ve -Gullstrand coordinates. By writing down the action whose imaginary part contributes to the tunneling effect, we have provided an estimate of the tunneling probability. Some comparison remarks are made in this regard.

\section{Acknowledgment}

We thank Rahul Ghosh for valuable discussions. We also thank the referee for constructive remarks. One of us (SS) thanks Shiv Nadar University for the grant of a research fellowship.

\section{Data availability statement}

All data supporting the findings of this study are included in the article.

 \end{document}